\documentclass[a4paper,12pt]{article}

\usepackage{amsmath,amssymb,graphics}
\usepackage[active]{srcltx}
\numberwithin{equation}{section}

\newcommand{\ii}{\mathrm{i}}
\newcommand{\na}{\nabla}
\newcommand{\dd}{\mathrm{d}}
\newcommand{\pd}{\partial}

\newcommand{\F}{\mathcal{F}}
\newcommand{\M}{\mathcal{M}}

\newcommand{\e}{\mathrm{e}}
\newcommand{\ket}[1]{\left|#1\right\rangle}
\newcommand{\bra}[1]{\left\langle #1\right|}

\newcommand{\tr}{\mathop{\mathrm{tr}}\nolimits}
\newcommand{\Tr}{\mathop{\mathrm{Tr}}\nolimits}

\newcommand{\A}{\mathfrak{A}}

\newcommand{\R}{\mathbb{R}}

\newcommand{\I}{\mathbb{I}}

\newcommand{\fc}{\check{\Phi}}
\newcommand{\ft}[2]{{\textstyle\frac{#1}{#2}}}
\newcommand{\qp}[1]{[\!\![ #1 ]\!\! ]}
\begin{document}

\title{On matrix models for anomalous dimensions of super Yang--Mills theory}%
\author{Stefano Bellucci$^{a,}$\thanks{e-mail: \texttt{bellucci@lnf.infn.it}}, \and
Corneliu Sochichiu$^{a,b,c,}$\thanks{On leave from~: Bogoliubov Lab. Theor. Phys., JINR,
141980 Dubna, Moscow Reg., RUSSIA; e-mail:
\texttt{sochichi@lnf.infn.it}}\\
$^a$ {\it INFN -- Laboratori Nazionali di Frascati,}\\
{\it Via E. Fermi 40, 00044 Frascati, Italy}\\
$^b$ {\it Max-Planck-Institut f\" ur Physik
(Werner-Heisenberg-Institut)}\\
{\it F\" ohringer Ring 6, D80805 M\" unchen}\\
$^c$ {\it Institutul de Fizic\u a Aplicat\u a A\c S,}\\
{\it str. Academiei, nr. 5, Chi\c{s}in\u{a}u, MD2028 Moldova}
}%

%

\maketitle
\begin{abstract}
 We consider the matrix model approach to the anomalous dimension matrix in $\mathcal{N}=4$
 super Yang--Mills theory. We construct the path integral
 representation for the anomalous dimension
 density matrix and analyze the resulting action. In particular, we consider the
 large $N$ limit, which results in a classical field theory. Since the same limit leads to
 spin chains, we propose to consider the former
 as an alternative description of the latter. We consider also the limit of small $N$,
 which corresponds to the restriction to the diagrams of maximal topological
 genus.
\end{abstract}
\maketitle
\section{Introduction}

The large $N$ approach \cite{'tHooft:1973jz} provides a description of the
quantized gauge models in terms of a topological expansion. This expansion is very
similar to the string perturbative expansion in terms of the geometrical genus. This
old idea to describe the strongly coupled gauge model as a string model has found
its so far best realization in the AdS/CFT conjecture
\cite{Maldacena:1998re,Gubser:1998bc}. As a true duality, the AdS/CFT correspondence
relates the weakly coupled theory on the one hand, with the strongly coupled on the
other. Once proven, this theory would become a powerful tool for the description
of both super Yang--Mills (SYM) theory and string theory. The same property,
however, prevents all possible proofs from being easy. The first approaches to the
problem relied heavily on the supersymmetry properties of the theories (see
\cite{Aharony:1999ti} for a review).

A substantial progress was achieved when it was realized that various limits of
the correspondence could be considered. In particular, a Penrose limit of the AdS
geometry \cite{Blau:2001ne,Blau:2002dy} was found to correspond to the pp-wave
geometry, in which the string theory is solvable \cite{Metsaev:2002re}. In SYM
theory this corresponds to operators with a large $R$-charge $J$. Various spinning
string solutions were found in
\cite{Frolov:2002av,Frolov:2003qc,Frolov:2003xy,Arutyunov:2003uj,Beisert:2003ea}
and the respective SYM sectors identified (see \cite{Tseytlin:2003ii} for a review
and a complete list of references).

On the other hand, the extensive study of the SYM anomalous dimensions revealed the
dilatation operator for one and higher loops \cite{Beisert:2003jj,Beisert:2004ry},
leading as well to the discovery of integrability in the planar limit
\cite{Minahan:2002ve,Beisert:2003yb}. Integrable structures discovered in SYM were
compared to those of string theory and a pretty good match was found
\cite{Arutyunov:2003rg,Arutyunov:2003za,Arutyunov:2004xy}.

At the same time, the nonplanar regime of SYM theory and AdS/CFT correspondence was
paid much less attention to. In the two-impurity sector of the BMN limit it was
shown \cite{Beisert:2002ff} that nonplanar corrections correspond to
splitting/joining of string-like configurations. The corresponding spin description
with the dynamical chain formation of the complete one-loop dilatation
operator was introduced in
\cite{Bellucci:2004ru,Bellucci:2004qx}. Such a nonplanar analysis was then further pursued
at the two- \cite{Bellucci:2004dv} and higher-loop \cite{Bellucci:2005cu} levels.

In this paper we are returning a little bit back and consider the anomalous
dimension operator of \cite{Beisert:2003jj} without going to the spin description.
As noted in \cite{Agarwal:2004cb} this operator can be regarded as the
Hamiltonian of a matrix model in the Schr\"{o}dinger picture\footnote{For an earlier
matrix model approach in BMN limit see \cite{deMelloKoch:2003pv}}. We adopt this
point of view and analyze the respective model. In particular, we propose a path
integral formula for the anomalous dimension density and analyze the large, as well
as the small $N$ limit of the path integral. The model is described in terms of a
gauge theory on a compact noncommutative space. The obtained model, however, is
different from the noncommutative Yang--Mills theory, due to a modified phase
structure. This modification can be regarded as an additional noncommutativity in the
space of gauge field configurations, modulo gauge symmetry. Next, we probe the
extremal limits of $N$. The limit of large $N$ in the spin description corresponds
to the integrable spin chain. In the case of the matrix model description, this is the
semiclassical limit, $1/N$ playing the role of the Planck constant $\hbar$. As a
result, physical quantities of the quantum spin chain, such as the partition
function or correlation functions, can be computed in terms of the matrix model, as
integrals over the moduli space of classical solutions.

A much less studied case (if studied at all) is the $N\to 0$ limit of a gauge theory. In the
topological expansion it corresponds to counting the contribution of diagrams with
the maximal topological genus. This limit cannot be reached directly within SYM
theory, since one cannot reach a vanishing value of $N$ continuously, because $N$, being
the rank of the gauge group, is restricted to integer values. The noncommutative
field theory description allows one to go beyond this limitation and extend the
description to arbitrary real values of $N$, which is compatible with the
$N\to 0$ limit. The
analysis of the star product expansion in this limit unveils its
strongly non local character. An alternative approach in terms of spin
bit model confirms this conclusion, the resulting model being given by
a nonlocal spin model.

The plan of the paper is as follows. In the next section we introduce the
anomalous one-loop dimension operator and find the corresponding matrix model, by
passing to the path integral description of the partition function or the Fourier
transform of the eigenvalue density. A gauge invariant matrix model of the reduced
Yang--Mills type emerges, as a result of taking into consideration the symmetries
of the system. Then, we find it suitable to represent the matrix model in terms of
a gauge model on a compact noncommutative space. We choose the simplest case of
the noncommutative torus, but other options like the fuzzy sphere are also
possible. Finally, we consider the large/small $N$ limit of the
obtained model and draw our conclusions.

\section{Matrix model for anomalous dimensions}
In this paper we consider the SU(2)-invariant sector of SYM operators which is
generated by all gauge invariant polynomials of two complex combinations of SYM
scalars. The one-loop anomalous dimension matrix for this sector was found in
\cite{Beisert:2003tq}. It can be written in a compact form as follows:
\begin{equation}\label{h2}
  H_{(2)}=-\frac{g_{\rm YM}^2}{16\pi^2}:\tr[\Phi^a,\Phi^b][\fc_a,\fc_b]:,
\end{equation}
where $\Phi^{a}$ $a=1,2$ are the mentioned complex combinations of the SU($N$)
scalar field: $\Phi^{1}=\phi_5+\ii\phi_6$ and $\Phi^{2}=\phi_1+\ii\phi_2$ and
checked letters correspond to the derivatives
\[
  \fc_{a,j}{}^i=\frac{\pd}{\pd\Phi^{a,}{}_i{}^j}.
\]

The operators under consideration are polynomials in $\Phi^a$ invariant under the
SU($N$) gauge transformation
\begin{equation}\label{gauge-transf}
  \Phi^{a}\to U^{-1}\Phi^{a}U,\qquad U\in \mathrm{SU}(N).
\end{equation}
Generally they can be imagined as a product of traces of products of $\Phi^a$'s.
An alternative parameterization can be introduced using a permutation group
element \cite{Bellucci:2004ru,Bellucci:2004qx}.

Therefore, the ``physical states'', i.e. the states on which the operator
\eqref{h2} is allowed to act, are given by gauge invariant polynomials of ``rising
operators'' $\Phi^a$. Formally, this corresponds to projecting onto the gauge
invariant sector of the Hilbert space, by imposing the following condition:
\begin{equation}\label{phys1}
  G\ket{\Psi}=0,
\end{equation}
where $G$ is the generator of gauge transformations
\begin{equation}
  G=-:[\Phi^a,\fc_a]:,
\end{equation}
i.e.
\begin{equation}
  \qp{\tr uG,\Phi^a}=[\Phi^a,u], \qquad  u\in \mathrm{su}(N),
\end{equation}
where the fat commutator $\qp{\cdot,\cdot}$ is the one defined over the Hilbert
space for which
\begin{equation}
  \qp{\fc_{a,j}{}^i,\Phi^{b,}{}_k{}^l}=\delta_a^b\delta^i_k\delta_j^l,
\end{equation}
and, in contrast to the usual one $[\cdot ,\cdot]$, it denotes the alternation in matrix matrix
products.

As noted in \cite{Agarwal:2004cb}, the operator \eqref{h2} can be regarded as the
quantized Hamiltonian in the Schr\"{o}dinger picture of a matrix model given by
the classical Hamiltonian\footnote{It coincides with the Hamiltonian of the matrix
model counting the combinatorial factors of Feynman diagrams proposed in
\cite{Beisert:2002ff}.}
\begin{equation}\label{h-cl}
  H_{\rm cl}=-\frac{g_{\rm YM}^2}{16\pi^2}\tr[X^a,X^b][\bar{X}_a,\bar{X}_b],
\end{equation}
the canonical Poisson bracket,
\begin{equation}\label{PB-cl}
  \{X^{a}{}_i{}^j,\bar{X}_b{}^k{}_l\}=\ii\delta^{a}{}_b\delta_i{}^k\delta^j_l,
\end{equation}
and the constraint
\begin{equation}\label{contr}
   G\equiv -[X^a,\bar{X}_a]\approx 0.
\end{equation}

The relation with the classical model given by \eqref{h-cl} and \eqref{PB-cl}, on the
one side, and the operator \eqref{h2} on the other, can be easily checked by
quantizing the former and going to the Schr\"{o}dinger picture. A little less
trivial approach is to go back and construct the path integral representation for
the operator \eqref{h2}, by considering the partition function
\begin{equation}\label{z-beta}
  Z_\tau=\Tr \e^{\ii\tau H_{(2)}}.
\end{equation}

The meaning of the partition function \eqref{z-beta} in the dual theory is clear:
under given boundary conditions it describes respective string amplitudes. A less
obvious fact is that it has an important meaning also in the original SYM theory.
In fact, up to a multiplicative factor, the partition function gives the
Hamiltonian $H_{(2)}$ eigenvalue density function, or the SYM anomalous dimension
distribution. Indeed,
\begin{equation}
  \rho(\lambda)=\frac{1}{2\pi}\int\dd\tau \tr\e^{\ii \tau(H_{(2)}-\lambda)}=
  \sum_k\delta(\lambda_k-\lambda),
\end{equation}
where the sum over the eigenvalues $\lambda_k$ should be understood in a broad
sense, including both summation over the discrete set and integration over the
continuous one.

Let us describe briefly the derivation of the path integral. As usual, one should
split the time interval $\tau$ in $L$ smaller pieces $\Delta=\tau/L$ and write the
exponential under the trace in \eqref{z-beta} as a product over these pieces
\begin{equation}\label{z-prod}
  Z_\tau=\Tr (\e^{-\Delta H_{(2)}})^L.
\end{equation}

Now we should employ the oscillator coherent states\footnote{See
\cite{Perelomov:book} for an exhaustive introduction to coherent states.}
\begin{multline}
  \ket{X}=
  \exp\tr(\bar{X}\phi-X\check{\phi})
  \cdot\exp\tr(\bar{Y}Z-X\check{Z})\ket{0}=\\
  \e^{-\tr \frac{1}{2}(\bar{X}X+\bar{Y}Y)}
  \e^{\bar{X}\phi+\bar{Y}Z}\ket{0}.
\end{multline}
Inserting the unity operator decomposition
\begin{equation}\label{unity-gni}
  \I=\int\dd^4X\ket{X}\bra{X}
\end{equation}
between each factor in \eqref{z-prod} and taking the limit $L\to\infty$, one gets
the partition function
\begin{equation}\label{pi}
  Z_{\tau}=\int[\dd X\,\dd \bar{X}]\,
  \e^{\ii S(X,\bar{X})},
\end{equation}
where the action $S(X,\bar{X})$ is given by
\begin{multline}\label{act-pi}
  S_t(X,\bar{X})=\\
  \int \dd t\,\left(\tr
  \frac{\ii}{2}(\bar{X}{}_a\dot{X}{}^a-\dot{\bar{X}}{}_aX^a)+
  \frac{g_{\rm YM}^2}{16\pi^2}\tr[X^a,X^b][\bar{X}_a,\bar{X}_b]\right),
\end{multline}
with $X^a$ and their Hermitian conjugate $\bar{X}_a$ being $N\times N$
time-dependent matrices in the adjoint representation of SU($N$).

The action \eqref{act-pi} is manifestly invariant, with respect to constant unitary
conjugation
\begin{align}\label{gt-pi}
  X^a&\to U^{-1}X^aU,\\
  \bar{X}_a&\to U^{-1}\bar{X}_aU,\qquad U\in \mathrm{SU}(N).
\end{align}
In fact, the full theory is invariant with respect to a bigger group of
time-dependent ($\dot{U}\neq 0$) gauge transformations. Indeed, an infinitesimal
transformation of the action \eqref{act-pi} with $U=\e^{u}\approx 1+u$ produces
the following term:
\begin{equation}
  \delta_{\rm gauge} S_t=-\tr\dot{u}[X^a,\bar{X}_a].
\end{equation}
The latter is proportional to the quantity $G(X)=-[X,\bar{X}]$, which for the physical
states is identically zero, since
\begin{equation}\label{constraint}
  G(X)=\bra{X}:[\Phi^a,\fc_a]:\ket{X}=[X^a,\bar{X}_a]\equiv 0.
\end{equation}
Therefore, the path integral \eqref{pi} can be regarded as the one corresponding to
the gauge invariant action\footnote{or the action for a system with the constraint
$G\approx 0$; in this case $\ii A_0$ is a Lagrange multiplier.}
\begin{equation}
  S=S_t+\frac{\ii}{2}\tr(\bar{X}_a[A_0,X^a]-[A_0,\bar{X}_a]X^a),
\end{equation}
where $A_0$ is the time (and unique) component of the SU($N$) gauge field, in the
temporal gauge $A_0=0$.

The gauge fixed non-invariant form of the action appeared, due to the gauge
non-invariance of the resolution of the unity operator in \eqref{unity-gni}. Indeed, the action of
the generator of the infinitesimal gauge transformation $\tr u G(\Phi,\fc)$,
$u\in$ SU($N$) on a coherent state $\ket{X}$ yields
\begin{equation}
  \tr u G\ket{X}=\ket{U^{-1}X U}-\ket{X}=\ket{X+[X,u]}-\ket{X}\neq 0.
\end{equation}
One can improve this situation by restricting the integration solely over the
physical coherent states satisfying \eqref{constraint}. This can be implemented by
introducing, in place of the unity, the projector to the physical sector
\begin{equation}\label{proj1}
  \Pi=\int\dd^4X\delta(G(X))\ket{X}\bra{X}=\Pi=\int\dd^4X\,\dd A_0\,
  \e^{\ii \tr A_0 G}\ket{X}\bra{X}.
\end{equation}

In spite of the manifest gauge invariance of the procedure, the expression for the
projector \eqref{proj1} cannot be considered yet completely satisfactory. In
fact, now there is ``too much'' gauge invariance, since both the integrand and the
measure are gauge invariant. Therefore, there is an extra dummy integration over
the gauge group SU($N$) in \eqref{proj1}. Normally, as SU($N$) is compact, this is not
a big trouble but, since we are inserting this integration an infinite number of
times, this could be a source of potential divergencies. This situation is
completely equivalent to that in any theory with gauge invariance and it is solved
in a similar way. Namely, one can explicitly break the gauge invariance by
introducing a gauge fixing term $F_{\rm gf}$. In this case the physical state
projector looks like
\begin{equation}\label{proj2}
  \Pi=\int\dd^4X\,\dd A_0\,\dd \chi\,\Delta_{\rm FP}(X) \e^{\ii \tr A_0G+
  \ii \tr\chi F_{\rm gf}}\ket{X}\bra{X},
\end{equation}
where $\chi$ is the Lagrange multiplier for the gauge fixing constraint $F_{\rm
gf}=0$ and $\Delta_{\rm FP}(X)$ is the famous Faddeev--Popov
determinant\footnote{For details regarding admissible gauge fixings,
Faddeev--Popov determinants and BRST invariance in a gauge theory, we refer the
reader to the classical reference \cite{Faddeev:1980be}.}, defined as
\[
  \Delta_{\rm FP}(X)=\int\dd U\, \delta(F_{\rm gf}(U^{-1}XU)).
\]

Now it is not difficult to see that, using the gauge fixed projector to the
physical sector \eqref{proj2} instead of unity, one gets the BRST invariant form
of the path integral,
\begin{equation}
  Z_\tau=\int[\dd X\,\dd A_0\,\dd\lambda\,\dd c\,\dd\bar{c}]\,
  \e^{\ii S_{\rm gi}(X,A)+\ii\tr\lambda F_{\rm gf}(X,A)+\ii \tr\bar{c}M_{\rm FP}(X,A)c},
\end{equation}
where
\begin{multline}\label{act-cl}
  S_{\rm gi}(X,A)=\\
  \int \dd t\,\left(\tr
  \frac{\ii}{2}(\bar{X}{}_a\nabla_0{X}{}^a-\nabla_0{\bar{X}}{}_aX^a)
  +\frac{g_{\rm YM}^2}{16\pi^2}\tr[X^a,X^b][\bar{X}_a,\bar{X}_b]\right)
\end{multline}
is the gauge invariant action, $\nabla_0 X=\pd_0 X+[A_0,X]$,
$\lambda$ is the Lagrange multiplier implementing the gauge fixing
condition $F_{\rm gf}=0$, and $M_{\rm FP}$ is the Faddeev--Popov
operator defined by
\begin{equation}
  \delta_{\rm gauge}F_{\rm gf}=M_{\rm FP}(X,A)u.
\end{equation}
Obviously, the gauge transformation of the gauge field $A_0$ is given by
\begin{equation}
  A_0\to U^{-1}A_0U+U^{-1}\dot{U}.
\end{equation}


\section{Noncommutative torus representation}

The gauge invariant classical action \eqref{act-cl} resembles a lot a
Yang--Mills-type model. Drawn by this, we will rewrite in this section the action
\eqref{act-cl} in terms of Yang--Mills theory on a two-dimensional noncommutative
torus. In fact, the choice of the two dimensional torus is not special, rather it is
dictated just by the simplicity of the space. In general, using the compact form
of the maps of noncommutative gauge theories considered in
\cite{Sochichiu:2000bg,Sochichiu:2000kz,Kiritsis:2002py} (see also
\cite{Sochichiu:2002jh}), one can pass among different theories, within the class of Morita
equivalent noncommutative spaces \cite{Schwarz:1998qj}.

The two dimensional noncommutative torus is defined by the ``coordinate
operators'' $U$ and $V$ subject to the following commutation relations:
\begin{equation}\label{nc-alg}
  UV=qVU.
\end{equation}
This algebra can be embedded into a usual Heisenberg algebra
\begin{equation}
  U=\e^{2\pi\ii x^1},\quad V=\e^{2\pi\ii x^2},\quad [x^1,x^2]=\ii\theta,
\end{equation}
with $q=\e^{4\pi^2\ii\theta}$. On the other hand, when $q$ is a $N$-th root of
unity: $q^N=1$, i.e. when $\theta=1/2\pi N$, the dimensionality of the irreducible
representation of the algebra is finite and equal to $N$. Indeed, $U$ and $V$ can
be represented in terms of the following $N\times N$ unitary matrices:
\begin{equation}
  U_{mn}=\delta_{m+1,n}, \qquad V_{mn}=\e^{2\pi\ii m/N}\delta_{mn},
\end{equation}
where no summation is assumed over repeating indices and indices are periodic in
$N$, $m+N\sim m$.

We leave to the reader the proof that any arbitrary $N\times N$ matrix $F$ can be
expressed as a Weyl ordered polynomial of degree up to $N-1$ in respectively $U$
and $V$
\begin{equation}\label{nc-torus1}
  F=\sum_{m=0}^{N-1}\sum_{n=0}^{N-1}f_{mn} W^{mn},
\end{equation}
where $W^{mn}$ is the Weyl ordered product of $V^m$ and $U^n$. In
terms of the Heisenberg algebra embedding, one has
\begin{equation}
  W^{mn}=\e^{2\pi m x^1+2\pi nx^2}.
\end{equation}

Based on eq.\eqref{nc-torus1}, one can construct a one-to-one map from $N\times N$
matrices to functions on the unit two-dimensional torus
\begin{equation}
  F\mapsto F(x,y)=\sum_{mn}f_{mn}\e^{2\pi\ii mx+2\pi\ii
  ny}~.
\end{equation}
The matrix product under this map is replaced by the noncommutative star product
\begin{equation}\label{star}
  F\cdot G\mapsto F*G(x,y)=F(x,y)\e^{\frac{\ii}{2\pi N}
  (\overleftarrow{\pd}_x\overrightarrow{\pd}_y-
  \overleftarrow{\pd}_y\overrightarrow{\pd}_x)}G(x,y),
\end{equation}
where the left/right arrow indicates that the derivative acts on
$F(x,y)$ or $G(x,y)$ respectively.

Some other useful properties are that (i) the trace of a matrix is given by the
integral over the torus of the corresponding function
\begin{equation}
  \tr F=N \int_{T^2}\dd x\,\dd y\, F(x,y)
\end{equation}
and (ii) commutators of a noncommutative torus function with $x$ and $y$
correspond to the derivative over, respectively, $y$ and $x$
\begin{equation}
  [x,F]\mapsto \ii \theta \pd_y F(x,y),\qquad [y,F]\mapsto -\ii\theta\pd_x F(x,y),
\end{equation}
which allow one to express the derivatives of a function in an algebraic way and,
\emph{viceversa}, to rewrite algebraic expressions as derivatives.

Using all of the above properties, one can rewrite the gauge invariant action of the
matrix model in terms of fields on the noncommutative torus
\begin{multline}\label{act-nct}
  S_{\rm gi}(X,A)=\\
  N\int \dd t\,\dd x\,\dd y\,\left(
  \frac{\ii}{2}(\bar{X}{}_a\nabla_0{X}{}^a-\nabla_0{\bar{X}}{}_aX^a)
  +\frac{g_{\rm
  YM}^2}{16\pi^2}[X^a,X^b]_*[\bar{X}_a,\bar{X}_b]_*\right),
\end{multline}
where $X^a$ and $\bar{X}_a$ are now functions on the torus and the
star-commutators are defined using the star product \eqref{star}
\begin{equation}
  [F,G]_*=F*G-G*F.
\end{equation}
The action \eqref{act-nct} is gauge invariant with respect to local time dependent
star-gauge transformations
\begin{align}
  X^a&\to U^{-1}*X^a*U\\
  \bar{X}_a&\to U^{-1}*\bar{X}_a*U\\
  A_{0}&\to U^{-1}*A_{0}*U+U^{-1}*\pd_0{U},
\end{align}
where $U\equiv U(x,y,t)$ is a local time dependent U(1) gauge transformation
\begin{equation}
  U^**U\equiv U^{-1}*U=1.
\end{equation}

If the time derivative term in the action \eqref{act-nct} were of the second
order, i.e. $\nabla_0\bar{X}\nabla_0 X$, rather than of the first one, we could
rewrite \eqref{act-nct} shifting the fields $X$, $\bar{X}$ and  up to total
derivative terms, in terms of a Yang--Mills type of action
\begin{equation}\label{action-YM}
  S_{\rm gi}=-\frac{16\pi^2N}{g_{\rm
  YM}^2}\int \dd^3  x\, \F^{\mu\nu}\bar{\F}_{\mu\nu},
\end{equation}
with the gauge field strength defined as
\begin{equation}
  \F_{\mu\nu}=\pd_\mu A_\nu-\pd_\nu A_\mu +[A_\mu,A_\nu]_*
\end{equation}
and the spatial part of the gauge field $A_a$, $a=1,2$ defined through the
relation
\begin{equation}\label{X}
  X^a=\frac{4\pi}{g_{\rm YM}}(\ii{\theta^{-1}\epsilon_{ab}x_b+A_a)},\qquad
  \theta=1/2\pi N,
\end{equation}
where $\epsilon_{ab}$ is the two dimensional antisymmetric tensor with the only
non-zero components $\epsilon_{xy}=-\epsilon_{yx}=1$. Eq. \eqref{X} gives the
splitting of the matrix field $X^a$ into the partial derivative and the gauge
field parts. The first order time derivative of the action makes it not only
impossible rewriting the action in terms of the Yang--Mills model, but it makes
the action non-invariant with respect to Lorentz boosts, apart from the fact that
this symmetry is broken by the noncommutativity.

The case at hand can be regarded as a sort of Landau limit of the Yang--Mills type
model, where the symplectic structure of the type $\dd p_i\wedge\dd q^i$ is replaced
with the ``noncommutative'' one, of the type $\theta^{-1}_{ij}\dd x^i\wedge\dd
x^j$.

\subsection*{String interpretation}
It is very tempting to relate the perturbative SYM anomalous dimension matrix with a
nonperturbative string dynamics given in terms of branes. Let us try to find the
meaning of the obtained matrix model in this context.

Let us consider the BFSS type matrix model describing the dynamics of $N$ zero-branes
\cite{Banks:1997vh}. It is given by the action
\begin{equation}\label{bfss}
  S_{\rm BFSS}=\int \dd
  t\,\tr\left(\ft12(\na_0X_i)^2+\ft{g}{4}[X_i,X_j]^2\right),
\end{equation}
where $g$ is the string coupling and $X_i$, $i=1,\dots,9$ are $N\times N$
Hermitian matrices. The eigenvalues of the matrices $X_i$ have the meaning of
0-brane coordinates. A modification of BFSS model describing the dynamics of
holomorphic branes in two-dimensional complex space will be formulated in terms of
sl(N) matrices rather than the su(N) ones of \eqref{bfss} which correspond to real
coordinates. The modified action in this case takes the form
\begin{equation}\label{compl}
  S_{\rm c}=\int \dd
  t\,\tr\left(\ft12\overline{\na_0X_a}\na_0X^a-\ft{g}{4}\overline{[X_a,X_b]}[X^a,X^b]\right),
\end{equation}
where $a,b=1,2$ and the bar stands for the Hermitian conjugate quantity. This
model is almost our matrix model \eqref{act-cl} except for the kinetic term.

This model is invariant, with respect to the time dependent SU(N) gauge transformations
\eqref{gt-pi}. This symmetry of the zero branes reflects the Chan--Patton gauge
invariance of open strings. It also means that zero branes are charged, with
respect to some SU(N) gauge field. The interaction with an external ``matrix
(electro-)magnetic field'' $\A_i$ can be introduced by adding to the action the
following term:
\begin{equation}
  \Delta S=\ii e\int\dd t\, \tr\A_i(X) \dot{X}_i,
\end{equation}
where $e$ is the unit charge of a zero brane. In particular a ``constant''
magnetic field is given by
\begin{equation}
  \A_i=\ft12 \mathfrak{F}_{ij}X_j.
\end{equation}
Taking the value of the magnetic field
 such that it respects the holomorphic
structure, i.e. $\mathfrak{F}_{ab}=\mathfrak{F}_{\bar{a}\bar{b}}=0$,
$\mathfrak{F}_{\bar{a}b}=f\delta_{a\bar{b}}$ will lead to the following
modification of the action \eqref{compl}:
\begin{equation}\label{compl2}
  S_{\rm cm}=\int \dd
  t\,\tr\left(\ft12|\na_0X^a|^2+\ii\ft{ef}{2}(\bar{X}_a\dot{X}^a-\dot{\bar{X}}_aX^a)
  -\ft{g}{4}|[X^a,X^b]|^2\right),
\end{equation}
Now, rescaling $X,\bar{X}\to\sqrt{ef}X,\sqrt{ef}\bar{X}$ and taking $ef\to\infty$,
one gets
\begin{equation}
  S_{\rm Landau}=\int \dd
  t\,\tr\left(\ft{\ii}{2}(\bar{X}_a\dot{X}^a-\dot{\bar{X}}_aX^a)
  -\ft{g_*}{4}|[X^a,X^b]|^2\right),
\end{equation}
where $g_*=g/(ef)^2$ is the modified string coupling, which can be put into
correspondence with the analogous factor in \eqref{act-cl}
\begin{equation}
  g_*=\frac{g}{(ef)^2}=\frac{g_{\rm YM}^2}{4\pi^2}.
\end{equation}
As it can be noted, the string coupling $g$ in this limit should be very large, in
order to keep $g_*$ fixed.

The limit which we described is similar to the one yielding the noncommutative
description for open strings \cite{Chu:1998qz,Seiberg:1999vs}. It is remarkable
that here one ends up with a model which is noncommutative in both moduli
space and space-time.
\section{Extremal cases}

In this section we discuss the extremal cases for the value of $N$. Since the rank
of the gauge group $N$ is a free parameter of the model, one may hope to get
simplifications, when it goes to some particular extremal values. The best studied
case is of course the planar $N\to\infty$ limit, where the SYM coupling $g_{\rm
YM}$ scales according to $\lambda_{\rm pl}=g_{\rm YM}^2 N=$ fixed. As we see
immediately below, this limit results also in a great simplification of our matrix
model.

On the SYM side, in the planar limit, only topologically trivial SYM Feynman
diagrams survive. The number of contributing diagrams is drastically reduced and
the instanton contribution is vanishing, which allows one to expect that the
perturbation theory is exact and analytic in $\lambda_{\rm pl}$. Via AdS/CFT
correspondence, the planar limit corresponds to taking the limit of free strings on
AdS$_5\times S^5$, while the expansion in powers of $1/N$ corresponds to the
topological expansion in the theory of interacting strings.

Generally, a contribution of a SYM Feynman diagram with $V_3$ triple vertices,
$V_4$ quadruple vertices and $H$ holes comes with a factor \cite{'tHooft:1973jz}
\[
  (g_{\rm YM}^2N)^FN^{2-2H},
\]
where $F=V_4+\ft12V_3$. For a fixed value of $F$, the maximal number of holes in the
diagram is bounded by $2H=F+2$, since there is no contribution with a negative
power of $N$ (for fixed $g_{\rm YM}$). Thus, at any loop level, which is controlled
by the power of $g_{\rm YM}$, the topological class of the diagram is bounded from
both below and above, the planar limit describing the lowest part of this
expansion. The natural question which can be addressed is whether there is an effective
theory describing the opposite limit of the expansion. Formally, this limit is
achieved when $N$ goes to zero, keeping $g_{\rm YM}$ fixed at the same time.

As $N$ is finite, this results in two different choices in the description of the
same model; therefore one may conjecture that these two limits result in dual
models.

Before going to the detailed description of the limits, let us make the following remark.
While there is no problem with achieving the planar limit $N\to\infty$, from the
point of view of SYM theory the range of the gauge group $N$ is always a positive
integer and the limit $N\to 0$ cannot be reached smoothly. The same remains true
for the dimensions of matrices in the matrix model description. Fortunately, the
noncommutative torus description allows one to overcome this handicap. Since
$N$ enters as the commutativity parameter, one can continue it to arbitrary real
analytical values. For an arbitrary real noncommutativity parameter, however, the
representation of the algebra of the noncommutative torus becomes infinite dimensional
and, in some sense, this limit is similar to the $N\to\infty$ limit.
\subsection{Planar limit ($N\to\infty$)}


Let us fix the 't Hooft coupling to be $\lambda_{\rm pl}=g_{\rm YM}^2N$ and make
the following rescaling of the fields:
\begin{equation}\label{resc}
  X\mapsto (2\pi/g_{\rm YM})X.
\end{equation}
The action \eqref{act-nct} then takes the following form:
\begin{equation}\label{act-pl}
  S_{\rm g.i.}=\frac{(2\pi N)^2}{\lambda_{\rm pl}}\int_{\R^1\times T^2}\dd^3x
  \left(\ii\bar{X}\nabla_0X+\ft14|[X,X]_*|^2\right),
\end{equation}
where we dropped the indices $a,b,\dots$ of the matrices $X_a,X_b,\dots$etc. The
integration is performed over time times the unit torus $0\leq x^{1,2}<0$.

In the limit $N\to\infty$ the noncommutativity parameter $\theta=1/2\pi N$
vanishes and the star product in the action \eqref{act-pl} can be approximated by
the leading terms in $1/N$
\begin{equation}
  A*B\equiv A\e^{\ii\theta\overleftarrow{\pd}\times\overrightarrow{\pd}}
  \approx AB+\ft{\ii}{4\pi N}\{A,B\},\qquad [A,B]_*\approx \ft{\ii}{2\pi N}\{A,B\}
\end{equation}
where $\{\cdot,\cdot\}$ denotes the Poisson bracket defined as
\begin{equation}
  \{A,B\}=\pd_x A\pd_y B-\pd_yA\pd_xB.
\end{equation}

Making another rescaling of the fields similar to $X\mapsto (1/2\pi N)X$, one
arrives to the following form of the action:
\begin{equation}\label{act-pl2}
  (2\pi N)^4S_{\rm pl}=\frac{(2\pi N)^4}{\lambda_{\rm pl}}\int_{\R^1\times T^2}\dd^3x
  \left(\ii\bar{X}\nabla_0X+\ft14|\{X,X\}|^2\right),
\end{equation}
where the fields $X$ are functions on the ordinary (commutative) torus. This action
describes a charged membrane in a strong magnetic field.

The dependence on $N$ is reduced to a diverging factor $(2\pi N)^4$ in front of
the action. This factor is analogous to the factor $1/\hbar$ in the standard
definition of the path integral
\[
  \int\e^{\frac{\ii}{\hbar}S}.
\]
Therefore, the limit $N\to\infty$ corresponds to the semiclassical limit $\hbar\to
0$ in ordinary quantum mechanics. In other words, the diverging factor in the
exponential of the path integral restricts it to the configurations with minimal
action, i.e. to the classical ones.

Indeed, in the large $N$ limit the path integral \eqref{pi} is reduced to the
following expression:
\begin{multline}
  Z_{\tau}=\\
  \int [\dd X]\e^{\ii (2\pi N)^4 S_{\rm pl}(X)}=
  \int\dd X_0[\dd X_\bot]\e^{\ii(2\pi N)^4 (S_{\rm pl}(X_0)+S''_{\rm
  pl}(X_0)X_\bot^2+\dots)}\\
  =\int \dd X_0\det{}'[S''(X_0)]\, \e^{\ii (2\pi N)^4 S_{\rm pl}(X_0)},
\end{multline}
where the integration in the last line is performed over the moduli space of
classical solutions with the measure $\dd X_0$. Thus, if a classical solution
continuously depend on $D_\M$ parameters $y_i$, $i=1,\dots,D_\M$, the measure $\dd
X_0$ can be expressed as
\begin{equation}
  \dd X_0=\prod_{i=1}^{D_\M}\dd y_i\sqrt{\det_{ij}\int\dd^3x\,
  \pd_i\bar{X}\pd_jX},
\end{equation}
where $\pd_i=\pd/\pd y_i$ are partial derivatives, with respect to the solution
parameters. As we expect, the moduli space of the solutions has more than just one
connected component, therefore the integration over continuous parameters should
be supplemented with the summation of the connected components. In this case the
classical action is constant on each connected component, while it may vary from
component to component.

The study of the structure of the moduli space of the solution of the system
\eqref{act-pl2} and the comparison with the results obtained, e.g. via Bethe Ansatz
in the spin chain approach, goes beyond the scope of the present work and we leave it for a
future research.

\subsection{Anti-planar limit ($N\to0$)}

As we discussed above, the limit in which Feynman diagrams with maximal
topological genus dominate, formally corresponds to taking a fixed small $g_{\rm YM}$ and $N\to
0$. The analytic extension for achieving this limit is obtained using the
noncommutative torus representation of the matrix model \eqref{act-cl}, with
$\theta=1/2\pi N$ as the noncommutativity parameter. The representation of the
noncommutative torus algebra depends in a complicated manner on whether $N$ is
rational or not. In what follows, we avoid these subtleties and just continue the
definition of the action \eqref{act-nct} to arbitrary values of $N$, using the fact
that it depends on $N$ only through the star product definition and as an overall
factor of the action, both allowing non-integer values.

In contrast to the planar case, the limit of small $N$ has two complicating
effects. The small overall factor of the action indicates that the integration in the
partition function, as $N$ goes to zero, is spread over arbitrary field
configurations, irrelevant for the value of their classical action. Since the domain
of the field values is non-compact, the path integral diverges in each point. The
situation is similar to the strong coupling limit of Yang--Mills theory. In the
latter case one can get finite answers evaluating the model on the lattice, where the
gauge fields are represented as compact group valued variables, in contrast to
noncompact algebra valued continuous fields. In this case one can compute the
partition function or some other correlation functions, in order to see e.g. that
they correspond
to a confined system (for details see \cite{Creutz:1984mg}). Lattice
discretizations of gauge models on noncommutative tori were considered a few years
ago, in connection with the twisted Eguchi--Kawai model\footnote{See
\cite{Ambjorn:2002uk} and the related references for a recent review.}.

A dufferent approach to the problem can be based on the
fact that the star product \eqref{star} can be equivalently written in a ``dual''
form
\begin{equation}\label{dual}
  A*B(x)=\frac{1}{\det(\pi\theta)}\int\dd z\dd y\,\e^{2\ii
  z^a\theta^{-1}_{ab}y^b}
  A(x+y)B(x+z).
\end{equation}
Indeed, the kernel acting on the product of $A$ and $B$ can be represented as the
following Gaussian type integral:
\begin{equation}\label{dual2}
  \frac{1}{\det(\pi\theta)}\int\dd z\dd y\, e^{2\ii z^a\theta^{-1}_{ab}y^b+z^a\pd_a+y^b\pd'_b}
  =\e^{\ii\theta^{ab}\pd_a\pd'_b},
\end{equation}
where $\pd$ acts only on $A(x)$, while the primed derivative $\pd'$ acts only on
$B(x)$. The formal manipulation of \eqref{dual2} is given a precise
meaning to, in terms of the Fourier modes of $A(x)$ and $B(x)$. Note,
however that the intaegration in variables $y$ and $z$ should be
performed over an infinite range: $-\infty <y,z<+\infty $, in order to
have the Gaussian integral. The infinite range of integration can be
split into the toric integration and summation over the widing modes
in the following way:
\begin{equation}
   y^a=n^a+\tilde{y}^a,\qquad z^a=m^a+\tilde{z}^a,
\end{equation}
where $n^a\in \mathbb{Z}$ and $m^a\in\mathbb{Z}$ are the winding modes
and the tilded variables $\tilde{y}^a\in [0,1)$, $\tilde{z}^a\in
  [0,1)$ are the toric variables. Then, the star product can be
takes the following form:
\begin{equation}
  A*B(x)=\int_{T^2\times T^2} \dd^2 \tilde{y}\dd^2\tilde{z}
  K(y,z;\theta) A(x+y)B(x+z),
\end{equation}
where the integration is now performed over the tori and the kernel
$K(y,z;\theta)$ is given by the sum over the winding modes
\begin{equation}\label{kern}
  K(y,z;\theta)=\frac{1}{\det(\pi\theta)}\sum_{m,n}\e^{2\ii
  (z^a+m^a)\theta^{-1}_{ab}(y^a+n^a)},
\end{equation}
and we dropped the tildas from the toric variables $y$ and $z$.

In order to evaluate the $N\to 0$ limit of the matrix model, it suffices to take
the expansion of the kernel \eqref{kern} in the powers of
$\theta^{-1}$. Thus, for the zero the winding mode one has
\begin{multline}\label{big-th}
  A*B(x)=\int_{T^2\times T^2}\dd^2z\dd^2y\,
  \e^{2\ii [z^a]\theta^{-1}_{ab}[y^b]}A(z+y)B(x+z)=\\
  \int_{T^@\times T^2}\dd^2z\dd^2y(1+2\ii[z^a]\theta^{-1}_{ab}[y^b])
  A(x+y)B(x+z)=\\
  \int\dd^2z
  B(z)\int\dd^2yB(y)+2\ii\theta^{-1}_{ab}\int\dd^2z[z^a-x^a]B(z)
  \int\dd^2y[y^b-x^b]A(a),
\end{multline}
where $[\dots]$ is the see-saw function, defined as
\begin{equation}
  [x]=x-n,\qquad \text{for } n\leq x<n+1,\qquad n\in\mathbb{Z},
\end{equation}
which takes into account the periodicity of the variables. The
last term in the expansion \eqref{big-th} can be rewritten, using the
properties of the see-saw function, in the following form:
\begin{equation}
  2\ii\theta^{-1}_{ab}\int\dd^2z ([z^a]-[x^a]+\epsilon^a(x,z))B(z)
  \int\dd^2z ([y^b]-[x^b]+\epsilon^b(x,y))A(y),
\end{equation}
where $\epsilon^a(x,y)$ is the step function
\begin{equation}
  \epsilon^a(x,y)=
  \begin{cases}
    1, & 0\leq y^a<x^a\\
    0, &\text{otherwise}.
  \end{cases}
\end{equation}

The result of the expansion of the kernel \eqref{kern}
leads to a rather unpleasant conclusion: even the leading terms of the
star product in this limit are highly nonlocal, containing terms which
are integrals of $A$, $B$, $y^aA$ and $z^aB$ over the torus, as
well as indefinite integrals of $A$ and $B$, to which it is difficult
to attribute any meaning.

\subsubsection{Spin bit approach}
A wa out of this apparently hopeless situation, in the attempt to
understand the antiplanar limit of the dilatation operator, is provided by the spin
bit approach \cite{Bellucci:2004ru,Bellucci:2004qx}.

As we discussed earlier, the limit $N\to0$ corresponds to the strongly
coupled limit of the matrix model \eqref{act-cl}. In the strongly coupled
regime the path integral formulation does not offer a big advantage
with respect to the ``operator pictuure'' of \eqref{h2} since one can
evaluate the path integral explicitly. One can start with the
operator picture form \eqref{h2} of the dilatation operator, in order to map it
to a spin system.

The map is constructed as follows
\cite{Bellucci:2004ru,Bellucci:2004qx}. Let us choose the vacuum of the
model, which satisfies the constraint \eqref{phys1} and is cancelled
by both $\check{\Phi}_a$, $a=1,2$. The physical states of the model
are created by acting by U$(N)$-invariant polynomials of
$\Phi^a$,\footnote{Here we are ignoring the issue related to the trace
  identities in Lie algebras, which is due to the fact that not all
  polynomials of traces are algebrically independent. Normally, the
  true Hilbert space of the model is the one factorised over such
  identities. As soon as we deal with the analytic continuation over $N$,
  ignoring the trace identities is equivalent to the analytic
  continuation from large $N$.}
\begin{equation}
  \ket{\Psi}=P(\Phi)\ket{\Omega},\quad \check{\Phi}_a \ket{\Omega}=0,\quad
  P(U^{-1}\Phi U)=P(\Phi).
\end{equation}

The problem of classification of gauge invariant states is thus
reduced to the problem of the classification of invariant polynomials
$P(\Phi)$. An invariant polynomial of the order $L$ is given by the
product of traces of $\Phi_{a_1}, \Phi_{a_2}\dots,\Phi_{a_L}$, where
$a_k=1,2$. This can be encoded in a state
$\ket{\{a_1,\dots,a_L\},\gamma}$ ``encoding'' the spin data described
by the labels $\{a_1,\dots,a_L\}$ and the chain structure data
described by the permutation $\gamma\in S_L$, where $S_L$ is the
permutation group of the labels $1,2,\dots,L$. In order to complete the
correspondence one should identify the ``physically identical'' states
\begin{equation}\label{symm}
  \ket{\{a_{\sigma_1},\dots,a_{\sigma_L}\},\sigma^{-1}\gamma\sigma}\sim
   \ket{\{a_1,\dots,a_L\},\gamma},
\end{equation}
where $\sigma\in S_L$ is an arbitrary permutation and $\sigma_k\equiv
\sigma(k)$. The equivalence \eqref{symm} reflects the invariance of
the physical state with respect to relabellings.

The polynomial which corresponds to the state
$\ket{\{a_1,\dots,a_L\},\gamma}$ is given by
\begin{equation}\label{P}
  P[\{a_{\sigma_1},\dots,a_{\sigma_L}\},\gamma](\Phi)=\Phi^{a_1}_{i_1
  i_{\gamma_1}}\Phi^{a_2}_{i_2i_{\gamma_2}}\dots
  \Phi^{a_L}_{i_L i_{\gamma_L}}.
\end{equation}
It is clear that, due to the fact that $\gamma$ is a permutation, each
matrix index appears in \eqref{P} exactly twice: once as the left and
once as the right index.

The dilatation operator in the above representation can be found by
the direct evaluation of \eqref{h2} on the polynomials of the type of
$P[\{a_{\sigma_1},\dots,a_{\sigma_L}\},\gamma]$. As a result we have
\begin{equation}\label{sp_h}
  H_{(2)}=\frac{g_{\rm YM}^2}{16\pi^2}\sum_{kl}
  H_{kl}(N\delta_{k\gamma_l}
  +\Sigma_{k\gamma_l}),
\end{equation}
where the spin part of the Hamiltonian is given by the two site
Hamiltonian of the Heisenberg spin chain
\begin{equation}
  H_{kl}=2(\I-P_{kl}),
\end{equation}
and the chain part is given by the chain splitting/joining operator
$\Sigma_{kl}$ defined as\footnote{Note a slight difference in notations with
  \cite{Bellucci:2004ru,Bellucci:2004qx}.}
\begin{equation}
  \Sigma_{kl}\ket{\{a_{\sigma_1},\dots,a_{\sigma_L}\},\gamma}=
   (1-\delta_{kl})\ket{\{a_{\sigma_1},\dots,a_{\sigma_L}\},\gamma\sigma_{kl}}
\end{equation}

As one can see, the $N$ dependence of the dilatation operator in the
spin form is extremely simple: $N$ appears only as the coupling to the
planar part of the Hamiltonian \eqref{sp_h}. Therefore, taking the
limit $N\to 0$ with $g_{\rm YM}^2$ fixed, results just in the
elimination of the plananr part of the dilatation operator! The
resulting expression reads
\begin{equation}\label{ap}
  H_{ap}=\frac{g_{\rm YM}^2}{16\pi^2}\sum_{kl}
  H_{kl}\Sigma_{k\gamma_l}.
\end{equation}

Clearly, in the model describing by the Hailtonian \eqref{ap}
there is no local structure: each spin bit is interacting equally
with any other spin bits, having no preferable neighbor, a situation
which is completely opposite to the planar limit.


\section{Discussion}

In this paper we considered the matrix interpretation of the SYM
anomalous dimension operator in the SU(2) sector of the theory. We constructed a
matrix path integral representation of the trace of the exponential of the
anomalous dimension operator, which in the dual theory has the meaning of the
partition function, while in the original SYM model it gives the Fourier transform
of the anomalous dimension density.

The matrix model we obtained has a potential part very similar to the one of BFSS
matrix model, the difference being the first order kinetic term in our case. Such a
term can be obtained effectively by placing the BFSS type matrix model in a strong
magnetic field. This class of models could be interpreted, from the physical
viewpoint, as
describing the dynamics of the zero branes in such a magnetic field. If this
interpretation is correct, our approach gives the relation between the Yang--Mills
coupling, string coupling and background magnetic field.

To the best of our knowledge, this type of models has not been studied in the literature
before, so they can serve as a topic for a future research, as we expect them to
have interesting properties.

We hope that the matrix model representation will be useful in the semiclassical study of
anomalous dimensions in the nonplanar sector, by analyzing the
corresponding solutions to the equations of motion. This analysis in
some cases is much simpler
than the diagonalization of the Hamiltonian in
\cite{Bellucci:2004ru,Bellucci:2004qx}. This seems to be an alternative/dual
approach to that of the sigma model description in the spirit of
\cite{Kruczenski:2003gt,Stefanski:2004cw,Bellucci:2004qr} which
includes nonplanar effects. (The noncommutative
target space seems to be a common feature in both approaches.) As we introduce the
noncommutative space parametrization for our model, this becomes the path integrals
over the space of noncommutative function, as obtained by the canonical
quantization \cite{Acatrinei:2002sb,Acatrinei:2001wa}.

The path integral and noncommutative field theory representation turns out to be useful
also for the analysis of various extremal limits, e.g. when the parameter $N$ is
either large or small. As for the large $N$ limit, the model corresponding to it is well
known: it is the integrable Heisenberg XXX$_{1/2}$ spin chain (or, better to say,
the direct sum of the spin chains corresponding to all lengths of the chains
$L$\footnote{In the notations of \cite{Bellucci:2004ru}.}). In our approach it
corresponds to the semiclassical limit of the matrix model. Thus, knowing all the
classical solutions to the obtained model, one can compute various quantities in
the quantum Heisenberg model.

Another extremal case we analyze is the limit of small $N$.
Unfortunately, the expansion in inverse powers of the noncommutative
parameter does not lead to any nice physical model. The problem is
caused by highly non-smooth and nonlocal limit of the star
product. There may exist a hope for the analysis using a regularized version
of the large $\theta$ star product. So far this appears technically
difficult.
On the other hand, this limit
is facilitated in the spin approach, if one neglects the trace identity
issue. The surprising result
is that the antiplanar limit corresponds to just the elimination of the
planar contribution from the dilatation operator. This appears to be
possible, since the only $N$ dependence of the model comes through the
trace of unity, which we make vanish.

Another point we would like to mention is the relation of our matrix model to the
matrix models describing various brane systems. There is a temptation to make such
an identification. In principle, our matrix model can be obtained as a limit of
holomorphic brane dynamics in a strong magnetic field. Perhaps there is another
possibility to find it, in the limit of fast rotating branes. In this context, it is
interesting if one meets there the situation with different noncommutative phases,
similar to the one which can be found in noncommutative quantum mechanics (see
\cite{Bellucci:2001xp,Bellucci:2002yh}; see also \cite{Bellucci:2004ak} for applications to
physical processes).

As a future development, beyond the already mentioned directions, it would be
interesting to extend the analysis to the whole SYM spectrum and
beyond one loop. In particular, it
would be interesting to apply the matrix model approach to the study of doubling
effects in the presence of fermions
\cite{Bellucci:2003qi,Danielsson:2003yc,Bellucci:2003hq,Bellucci:2004gc,Bellucci:2004qx}. (Let us
note that the doubling problem in the context of matrix models was already
addressed in \cite{Kitsunezaki:1997iu,Sochichiu:2000fs,Dai:2003ak}).

\subsection*{Acknowledgements}
We thank Pierre-Yves Casteill and Francisco Morales for discussions which served
as a source of inspiration for this work.

This work was partially supported by NATO Collaborative Linkage Grant PST.CLG.
97938, INTAS-00-00254 grant, RF Presidential grants MD-252.2003.02,
NS-1252.2003.2, INTAS grant 03-51-6346, INTAS grant 00-00262, RFBR-DFG
grant 436 RYS 113/669/0-2, RFBR
grant 03-02-16193, the European Community's Human Potential Programme under
contract HPRN-CT-2000-00131 Quantum Spacetime
and the Marie Curie Research Training Network
under contract MRTN-CT-2004-005104 Forces Universe.

\bibliographystyle{utphys}
\bibliography{db1}

\providecommand{\href}[2]{#2}\begingroup\raggedright\begin{thebibliography}{10}

\bibitem{'tHooft:1973jz}
G.~'t~Hooft, ``A planar diagram theory for strong interactions,'' {\em Nucl.
  Phys.} {\bf B72} (1974)
461.

\bibitem{Maldacena:1998re}
J.~M. Maldacena, ``The large {$N$} limit of superconformal field theories and
  supergravity,'' {\em Adv. Theor. Math. Phys.} {\bf 2} (1998) 231--252,
\href{http://www.arXiv.org/abs/hep-th/9711200}{{\tt hep-th/9711200}}.

\bibitem{Gubser:1998bc}
S.~S. Gubser, I.~R. Klebanov, and A.~M. Polyakov, ``Gauge theory correlators
  from non-critical string theory,'' {\em Phys. Lett.} {\bf B428} (1998)
  105--114,
\href{http://www.arXiv.org/abs/hep-th/9802109}{{\tt hep-th/9802109}}.

\bibitem{Aharony:1999ti}
O.~Aharony, S.~S. Gubser, J.~M. Maldacena, H.~Ooguri, and Y.~Oz, ``Large {$N$}
  field theories, string theory and gravity,'' {\em Phys. Rept.} {\bf 323}
  (2000) 183--386,
\href{http://www.arXiv.org/abs/hep-th/9905111}{{\tt hep-th/9905111}}.

\bibitem{Blau:2001ne}
M.~Blau, J.~Figueroa-O'Farrill, C.~Hull, and G.~Papadopoulos, ``A new maximally
  supersymmetric background of {IIB} superstring theory,'' {\em JHEP} {\bf 01}
  (2002) 047,
\href{http://www.arXiv.org/abs/hep-th/0110242}{{\tt hep-th/0110242}}.

\bibitem{Blau:2002dy}
M.~Blau, J.~Figueroa-O'Farrill, C.~Hull, and G.~Papadopoulos, ``Penrose limits
  and maximal supersymmetry,'' {\em Class. Quant. Grav.} {\bf 19} (2002)
  L87--L95,
\href{http://www.arXiv.org/abs/hep-th/0201081}{{\tt hep-th/0201081}}.

\bibitem{Metsaev:2002re}
R.~R. Metsaev and A.~A. Tseytlin, ``Exactly solvable model of superstring in
  plane wave {Ramond}--{Ramond} background,'' {\em Phys. Rev.} {\bf D65} (2002)
  126004,
\href{http://www.arXiv.org/abs/hep-th/0202109}{{\tt hep-th/0202109}}.

\bibitem{Frolov:2002av}
S.~Frolov and A.~A. Tseytlin, ``Semiclassical quantization of rotating
  superstring in {AdS$_5\times S^5$},'' {\em JHEP} {\bf 06} (2002) 007,
\href{http://www.arXiv.org/abs/hep-th/0204226}{{\tt hep-th/0204226}}.

\bibitem{Frolov:2003qc}
S.~Frolov and A.~A. Tseytlin, ``Multi-spin string solutions in {AdS$_5\times
  S^5$},'' {\em Nucl. Phys.} {\bf B668} (2003) 77--110,
\href{http://www.arXiv.org/abs/hep-th/0304255}{{\tt hep-th/0304255}}.

\bibitem{Frolov:2003xy}
S.~Frolov and A.~A. Tseytlin, ``Rotating string solutions: {AdS/CFT} duality in
  non-supersymmetric sectors,'' {\em Phys. Lett.} {\bf B570} (2003) 96--104,
\href{http://www.arXiv.org/abs/hep-th/0306143}{{\tt hep-th/0306143}}.

\bibitem{Arutyunov:2003uj}
G.~Arutyunov, S.~Frolov, J.~Russo, and A.~A. Tseytlin, ``Spinning strings in
  {AdS$_5\times S^5$} and integrable systems,'' {\em Nucl. Phys.} {\bf B671}
  (2003) 3--50,
\href{http://www.arXiv.org/abs/hep-th/0307191}{{\tt hep-th/0307191}}.

\bibitem{Beisert:2003ea}
N.~Beisert, S.~Frolov, M.~Staudacher, and A.~A. Tseytlin, ``Precision
  spectroscopy of {AdS/CFT},'' {\em JHEP} {\bf 10} (2003) 037,
\href{http://www.arXiv.org/abs/hep-th/0308117}{{\tt hep-th/0308117}}.

\bibitem{Tseytlin:2003ii}
A.~A. Tseytlin, ``Spinning strings and {AdS/CFT} duality,''
\href{http://www.arXiv.org/abs/hep-th/0311139}{{\tt hep-th/0311139}}.

\bibitem{Beisert:2003jj}
N.~Beisert, ``The complete one-loop dilatation operator of {$\mathcal{N} = 4$}
  super {Yang--Mills} theory,'' {\em Nucl. Phys.} {\bf B676} (2004) 3--42,
\href{http://www.arXiv.org/abs/hep-th/0307015}{{\tt hep-th/0307015}}.

\bibitem{Beisert:2004ry}
N.~Beisert, ``The dilatation operator of {$\mathcal{N} = 4$} super
  {Yang--Mills} theory and integrability,''
\href{http://www.arXiv.org/abs/hep-th/0407277}{{\tt hep-th/0407277}}.

\bibitem{Minahan:2002ve}
J.~A. Minahan and K.~Zarembo, ``The {Bethe--Ansatz} for {$\mathcal{N} = 4$}
  super {Yang--Mills},''
\href{http://www.arXiv.org/abs/hep-th/0212208}{{\tt hep-th/0212208}}.

\bibitem{Beisert:2003yb}
N.~Beisert and M.~Staudacher, ``The {$\mathcal{N} = 4$} integrable super spin
  chain,'' {\em Nucl. Phys.} {\bf B670} (2003) 439--463,
\href{http://www.arXiv.org/abs/hep-th/0307042}{{\tt hep-th/0307042}}.

\bibitem{Arutyunov:2003rg}
G.~Arutyunov and M.~Staudacher, ``Matching higher conserved charges for strings
  and spins,'' {\em JHEP} {\bf 03} (2004) 004,
\href{http://www.arXiv.org/abs/hep-th/0310182}{{\tt hep-th/0310182}}.

\bibitem{Arutyunov:2003za}
G.~Arutyunov, J.~Russo, and A.~A. Tseytlin, ``Spinning strings in {$\mathcal{N}
  = 4$}: New integrable system relations,'' {\em Phys. Rev.} {\bf D69} (2004)
  086009,
\href{http://www.arXiv.org/abs/hep-th/0311004}{{\tt hep-th/0311004}}.

\bibitem{Arutyunov:2004xy}
G.~Arutyunov and M.~Staudacher, ``Two-loop commuting charges and the string /
  gauge duality,''
\href{http://www.arXiv.org/abs/hep-th/0403077}{{\tt hep-th/0403077}}.

\bibitem{Beisert:2002ff}
N.~Beisert, C.~Kristjansen, J.~Plefka, and M.~Staudacher, ``{BMN} gauge theory
  as a quantum mechanical system,''
\href{http://www.arXiv.org/abs/hep-th/0212269}{{\tt hep-th/0212269}}.

\bibitem{Bellucci:2004ru}
S.~Bellucci, P.~Y. Casteill, J.~F. Morales, and C.~Sochichiu, ``Spin bit models
  from non-planar {$\mathcal{N} = 4$ SYM},'' {\em Nucl. Phys.} {\bf B699}
  (2004) 151--173,
\href{http://www.arXiv.org/abs/hep-th/0404066}{{\tt hep-th/0404066}}.

\bibitem{Bellucci:2004qx}
S.~Bellucci, P.~Y. Casteill, J.~F. Morales, and C.~Sochichiu, ``Chaining spins
  from (super){Yang--Mills},''
\href{http://www.arXiv.org/abs/hep-th/0408102}{{\tt hep-th/0408102}}.

\bibitem{Bellucci:2004dv}
S.~Bellucci, P.~Y. Casteill, A. Marrani, and C.~Sochichiu, ``Spin bits at two loops,''
{\em Phys. Lett.} {\bf B607} (2005) 180--187,
\href{http://www.arXiv.org/abs/hep-th/0411261}{{\tt hep-th/0411261}}.

\bibitem{Bellucci:2005cu}
S.~Bellucci, and A. Marrani, ``Non-planar spin bits beyond two loops,''
\href{http://www.arXiv.org/abs/hep-th/0505106}{{\tt hep-th/0505106}}.

\bibitem{Agarwal:2004cb}
A.~Agarwal, and S.~G. Rajeev, ``The dilatation operator of $\mathcal{N} = 4$
  {SYM} and classical limits of spin chains and matrix models,''
\href{http://www.arXiv.org/abs/hep-th/0405116}{{\tt hep-th/0405116}}.

\bibitem{deMelloKoch:2003pv}
R.~de~Mello~Koch, A.~Donos, A.~Jevicki, and J.~P. Rodrigues, ``Derivation of
  string field theory from the large {$N$} {BMN} limit,'' {\em Phys. Rev.} {\bf
  D68} (2003) 065012,
\href{http://www.arXiv.org/abs/hep-th/0305042}{{\tt hep-th/0305042}}.

\bibitem{Beisert:2003tq}
N.~Beisert, C.~Kristjansen, and M.~Staudacher, ``The dilatation operator of
  $\mathcal{N} = 4$ super {Yang--Mills} theory,'' {\em Nucl. Phys.} {\bf B664}
  (2003) 131--184,
\href{http://www.arXiv.org/abs/hep-th/0303060}{{\tt hep-th/0303060}}.

\bibitem{Perelomov:book}
A.~Perelomov, {\em Generalized Coherent States and their Applications}.
\newblock Springer-Verlag, Berlin, 1986.

\bibitem{Faddeev:1980be}
L.~D. Faddeev, and A.~A. Slavnov, ``Gauge fields. {I}ntroduction to quantum
  theory,'' {\em Front. Phys.} {\bf 50} (1980)
1--232.

\bibitem{Sochichiu:2000bg}
C.~Sochichiu, ``On the equivalence of noncommutative models in various
  dimensions,'' {\em JHEP} {\bf 08} (2000) 048,
\href{http://www.arXiv.org/abs/hep-th/0007127}{{\tt hep-th/0007127}}.

\bibitem{Sochichiu:2000kz}
C.~Sochichiu, ``Exercising in {K}-theory: Brane condensation without tachyon,''
\href{http://www.arXiv.org/abs/hep-th/0012262}{{\tt hep-th/0012262}}.

\bibitem{Kiritsis:2002py}
E.~Kiritsis, and C.~Sochichiu, ``Duality in non-commutative gauge theories as a
  non-perturbative {Seiberg--Witten} map,''
\href{http://www.arXiv.org/abs/hep-th/0202065}{{\tt hep-th/0202065}}.

\bibitem{Sochichiu:2002jh}
C.~Sochichiu, ``Gauge invariance and noncommutativity,''
\href{http://www.arXiv.org/abs/hep-th/0202014}{{\tt hep-th/0202014}}.

\bibitem{Schwarz:1998qj}
A.~Schwarz, ``Morita equivalence and duality,'' {\em Nucl. Phys.} {\bf B534}
  (1998) 720--738,
\href{http://www.arXiv.org/abs/hep-th/9805034}{{\tt hep-th/9805034}}.

\bibitem{Banks:1997vh}
T.~Banks, W.~Fischler, S.~H. Shenker, and L.~Susskind, ``M theory as a matrix
  model: A conjecture,'' {\em Phys. Rev.} {\bf D55} (1997) 5112--5128,
\href{http://www.arXiv.org/abs/hep-th/9610043}{{\tt hep-th/9610043}}.

\bibitem{Chu:1998qz}
C.-S. Chu, and P.-M. Ho, ``Noncommutative open string and {D}-brane,'' {\em
  Nucl. Phys.} {\bf B550} (1999) 151--168,
\href{http://www.arXiv.org/abs/hep-th/9812219}{{\tt hep-th/9812219}}.

\bibitem{Seiberg:1999vs}
N.~Seiberg, and E.~Witten, ``String theory and noncommutative geometry,'' {\em
  JHEP} {\bf 09} (1999) 032,
\href{http://www.arXiv.org/abs/hep-th/9908142}{{\tt hep-th/9908142}}.

\bibitem{Creutz:1984mg}
M.~Creutz, {\em Quarks, gluons and lattices}.
\newblock Cambridge Monographs on Mathematical Physics. Cambridge Univ. Press,
  1983.

\bibitem{Ambjorn:2002uk}
J.~Ambjorn, ``Strings, quantum gravity and noncommutative geometry on the
  lattice,'' {\em Grav. Cosmol.} {\bf 8} (2002)
144--150.

\bibitem{Kruczenski:2003gt}
M.~Kruczenski, ``Spin chains and string theory,''
\href{http://www.arXiv.org/abs/hep-th/0311203}{{\tt hep-th/0311203}}.

\bibitem{Stefanski:2004cw}
J.~Stefanski, B. and A.~A. Tseytlin, ``Large spin limits of {AdS/CFT} and
  generalized {Landau}--{Lifshitz} equations,'' {\em JHEP} {\bf 05} (2004) 042,
\href{http://www.arXiv.org/abs/hep-th/0404133}{{\tt hep-th/0404133}}.

\bibitem{Bellucci:2004qr}
S.~Bellucci, P.~Y. Casteill, J.~F. Morales, and C.~Sochichiu, ``sl(2) spin
  chain and spinning strings on {AdS$_5\times S^5$},'' {\em Nucl. Phys.} {\bf
  B707} (2005) 303--320,
\href{http://www.arXiv.org/abs/hep-th/0409086}{{\tt hep-th/0409086}}.

\bibitem{Acatrinei:2002sb}
C.~Acatrinei, ``Canonical quantization of noncommutative field theory,'' {\em
  Phys. Rev.} {\bf D67} (2003) 045020,
\href{http://www.arXiv.org/abs/hep-th/0204197}{{\tt hep-th/0204197}}.

\bibitem{Acatrinei:2001wa}
C.~Acatrinei, ``Path integral formulation of noncommutative quantum
  mechanics,'' {\em JHEP} {\bf 09} (2001) 007,
\href{http://www.arXiv.org/abs/hep-th/0107078}{{\tt hep-th/0107078}}.

\bibitem{Bellucci:2001xp}
S.~Bellucci, A.~Nersessian, and C.~Sochichiu, ``Two phases of the
  non-commutative quantum mechanics,'' {\em Phys. Lett.} {\bf B522} (2001)
  345--349,
\href{http://www.arXiv.org/abs/hep-th/0106138}{{\tt hep-th/0106138}}.

\bibitem{Bellucci:2002yh}
S.~Bellucci, and A.~Nersessian, ``Phases in noncommutative quantum mechanics on
  (pseudo)sphere,'' {\em Phys. Lett.} {\bf B542} (2002) 295--300,
\href{http://www.arXiv.org/abs/hep-th/0205024}{{\tt hep-th/0205024}}.

\bibitem{Bellucci:2004ak}
S.~Bellucci, and A.~Yeranyan, ``Noncommutative quantum scattering in a central field,''
{\em Phys. Lett.} {\bf B609} (2005) 418--423,
\href{http://www.arXiv.org/abs/hep-th/0412305}{{\tt hep-th/0412305}}.

\bibitem{Bellucci:2003qi}
S.~Bellucci, and C.~Sochichiu, ``Fermion doubling and {BMN} correspondence,''
  {\em Phys. Lett.} {\bf B564} (2003) 115--122,
\href{http://www.arXiv.org/abs/hep-th/0302104}{{\tt hep-th/0302104}}.

\bibitem{Danielsson:2003yc}
U.~Danielsson, F.~Kristiansson, M.~Lubcke, and K.~Zarembo, ``String bits
  without doubling,'' {\em JHEP} {\bf 10} (2003) 026,
\href{http://www.arXiv.org/abs/hep-th/0306147}{{\tt hep-th/0306147}}.

\bibitem{Bellucci:2003hq}
S.~Bellucci, and C.~Sochichiu, ``Can string bits be supersymmetric?,'' {\em
  Phys. Lett.} {\bf B571} (2003) 92--96,
\href{http://www.arXiv.org/abs/hep-th/0307253}{{\tt hep-th/0307253}}.

\bibitem{Bellucci:2004gc}
S.~Bellucci, and C.~Sochichiu, ``On the dynamics of BMN operators of finite size and the
model of string  bits,''
\href{http://www.arXiv.org/abs/hep-th/0404143}{{\tt hep-th/0404143}}.

\bibitem{Kitsunezaki:1997iu}
N.~Kitsunezaki, and J.~Nishimura, ``Unitary {IIB} matrix model and the dynamical
  generation of the space time,'' {\em Nucl. Phys.} {\bf B526} (1998) 351--377,
\href{http://www.arXiv.org/abs/hep-th/9707162}{{\tt hep-th/9707162}}.

\bibitem{Sochichiu:2000fs}
C.~Sochichiu, ``Matrix models: Fermion doubling vs. anomaly,'' {\em Phys.
  Lett.} {\bf B485} (2000) 202--207,
\href{http://www.arXiv.org/abs/hep-th/0005156}{{\tt hep-th/0005156}}.

\bibitem{Dai:2003ak}
J.~Dai, and Y.-S. Wu, ``Quiver mechanics for deconstructed matrix string,'' {\em
  Phys. Lett.} {\bf B576} (2003) 209--218,
\href{http://www.arXiv.org/abs/hep-th/0306216}{{\tt hep-th/0306216}}.

\end{thebibliography}\endgroup

\end{document}